\pdfoutput=1
\documentclass[12pt]{article}

\usepackage{graphicx,psfrag,epsf,color}
\usepackage{amsmath,amssymb,amsfonts}
\usepackage{array}
\usepackage{cite}
\usepackage{multirow}
\usepackage{rotating}
\usepackage{slashed}

\newcounter{MBQ}

\newcommand{\be}{\begin{equation}}
\newcommand{\ee}{\end{equation}}
\newcommand{\bea}{\begin{eqnarray}}
\newcommand{\eea}{\end{eqnarray}}
\newcommand{\bi}{\begin{itemize}}
\newcommand{\ei}{\end{itemize}}
\newcommand{\ben}{\begin{enumerate}}
\newcommand{\een}{\end{enumerate}}
\newcommand{\bt}{\begin{tabular}}
\newcommand{\et}{\end{tabular}}

\newcommand{\Eres}{E^\gamma_{\rm res}}
\newcommand{\mchi}{m_\chi}
\newcommand{\la}{\lambda}
\newcommand{\np}{n_+}
\newcommand{\nm}{n_-}

\begin{document}
\allowdisplaybreaks

\begin{titlepage}

\begin{flushright}
{\small
TUM-HEP-1139/18\\
arXiv:1805.07367\\[0.0cm]
May 01, 2020
}
\end{flushright}

\vskip1cm
\begin{center}
{\Large \bf Energetic $\gamma$-rays from TeV scale dark matter\\[0.1cm] 
annihilation resummed}\\[0.2cm]
\end{center}

\vspace{0.5cm}
\begin{center}
{\sc M.~Beneke$^{a}$, A.~Broggio$^{a}$, C.~Hasner$^{a}$,} \\ 
and  {\sc M.~Vollmann$^{a}$}\\[6mm]
{\it ${}^a$Physik Department T31,\\
James-Franck-Stra\ss e~1, 
Technische Universit\"at M\"unchen,\\
D--85748 Garching, Germany}
\\[0.3cm]
\end{center}

\vspace{0.6cm}
\begin{abstract}
\vskip0.2cm\noindent
The annihilation cross section of TeV scale dark matter particles 
$\chi^0$ with electroweak charges into photons is affected by large quantum 
corrections due to Sudakov logarithms and the Sommerfeld effect. 
We calculate the semi-inclusive photon energy spectrum in 
$\chi^0\chi^0\to \gamma+X$ in the vicinity of the maximal photon 
energy $E_\gamma = m_\chi$ with NLL' accuracy in an all-order 
summation of the electroweak perturbative expansion adopting the 
pure wino model. This results in the most precise theoretical 
prediction of the annihilation rate for $\gamma$-ray telescopes
with photon energy resolution of parametric order $m_W^2/m_\chi$ for 
photons with TeV energies.
\end{abstract}
\end{titlepage}

\section{Introduction}
\label{sec:introduction}

Within the large variety of dark matter (DM) candidates a weakly interacting 
particle with mass $\mchi$ in the 100~GeV to 10~TeV range (WIMP) stands out 
due to its conceptual minimality and its relation to the electroweak scale. 
Although loop suppressed, the pair annihilation of WIMPs into two photons or 
a photon and a $Z$ boson often provides a distinctive signature in the form 
of a monochromatic component of high-energy cosmic $\gamma$-rays. The 
strongest limit on such $\gamma$-line signals is currently set by the 
H.E.S.S. experiment \cite{Abdallah:2018qtu}. This limit is expected to be 
improved by an order of magnitude by the 
Cherenkov Telescope Array (CTA) \cite{Consortium:2010bc} 
under construction, which will severely constrain generic WIMP models. This 
motivates a thorough theoretical investigation of the basic pair annihilation 
cross section given a particular model.

It is well-known that TeV-scale DM annihilation is not accurately described 
by the leading-order annihilation rate, but is modified by the  
Sommerfeld effect \cite{Hisano:2004ds} generated by the electroweak Yukawa 
force on the DM particles prior to their annihilation. In terms of 
Feynman diagrams the Sommerfeld effect corresponds to ladder diagrams with 
$W$, $Z$, photon and Higgs boson exchange, which contribute 
${\cal O}((\mchi \alpha_2/ m_W)^n)$ at order $n$, where $m_W$ is the 
$W$ boson mass and $\alpha_2$ the SU(2) gauge coupling. For annihilation 
rates to exclusive final states, in addition to the Sommerfeld effect large 
logarithmically enhanced quantum corrections of 
${\cal O}((\alpha_2\ln^2 (\mchi/m_W))^n)$, also known as electroweak Sudakov 
logarithms, arise \cite{Hryczuk:2011vi} from the restriction on the emission 
of soft 
radiation.\footnote{Since the DM particles carry electroweak charges, Sudakov 
logarithms remain even in the inclusive annihilation rate relevant to relic 
density calculations, but the effect is suppressed due to the presence of 
coannihilation of the nearly degenerate full electroweak multiplet.} 
Electroweak Sudakov logarithms in DM annihilation into photons have been 
identified as a potential source of large corrections and resummed to all 
orders in perturbation theory in previous 
work \cite{Baumgart:2014vma,Bauer:2014ula,Baumgart:2015bpa,Ovanesyan:2014fwa,Ovanesyan:2016vkk} 
in models with an electroweak triplet scalar or fermionic DM particle.

In this letter we revisit this question starting from the observation that 
telescopes do not measure two photons from a single decay in coincidence. 
Rather, the observable is the {\em semi-inclusive} single-photon 
energy spectrum $\gamma+X$, where $X$ denotes the unidentified other final 
state particles. This spectrum is composed of two spectral line features 
related to the $\gamma\gamma$ and $\gamma Z$ final states and a continuum 
from multi-body final states. While 
the leading term in the expansion in $\alpha_2$ is indeed exclusively from 
the $\gamma\gamma$ and $\gamma Z$ final state, the logarithmically 
enhanced quantum 
corrections are different for the exclusive and the semi-inclusive 
measurement. If $\Eres\ll \mchi$ denotes the energy resolution of the 
instrument for photon energies $E_\gamma\approx \mchi$, the maximal 
invariant mass of the unobserved final state $X$ is $m_X^2 = 4 \mchi\Eres 
\ll \mchi^2$. Since $X$ must balance the large momentum of the observed 
photon, $X$ is forced to have a `jet-like' structure. The kinematic 
situation involving non-relativistic heavy particles in the initial 
state and energetic, small-invariant mass objects in the final state 
is naturally described by a combination of non-relativistic and 
soft-collinear effective field theory, similar to the QCD treatment 
of the `inverse' situation of hadronic production of two heavy 
particles \cite{Beneke:2010da}.
 
In the following sections we first summarize briefly the basic elements of 
an effective field theory (EFT) treatment of the single-inclusive photon 
spectrum $d(\sigma v)/dE_\gamma$ in DM pair annihilation near the kinematic 
endpoint. For the DM model 
we refer to the widely discussed pure wino model, which features an 
electroweak triplet whose electrically neutral component is the DM particle, 
although all results except the one for the soft function and the 
Sommerfeld factor apply more generally to DM particles in an arbitrary 
isospin-$j$ multiplet.  We then present and discuss our result for the 
all-order resummed spectrum including both the Sommerfeld and Sudakov 
corrections. A more detailed exposition of the formalism as well as 
extensions will be reported in a longer article. While this work was being 
finalized, a similar EFT 
calculation of the endpoint of the $\gamma+X$ spectrum has 
appeared \cite{Baumgart:2017nsr}. The present EFT formulation 
refers to a finer photon energy resolution, but includes the one-loop 
corrections to all  matching coefficients, soft and jet functions thus 
achieving NLL' rather than LL accuracy for the observable in question.


\section{The resummed energy spectrum}
\label{sec:theory}

We add to the Standard Model (SM) Lagrangian a fermionic multiplet $\chi$ 
(which can be of Majorana or Dirac type) in an arbitrary isospin-$j$ 
representation of the electroweak (EW) SU(2) gauge group. For the Majorana 
case, only integer $j$ are allowed, while for the Dirac case also 
half-integer $j$ are possible. In the calculations below we assume zero 
hypercharge ($Y=0$), in which case only integer $j$ multiplets provide a
realistic DM particle as the electrically neutral member $\chi^0$ 
of the $2j+1$ dimensional multiplet. The Lagrangian is 
\begin{equation}
\mathcal{L} = \mathcal{L}_{\text{SM}} 
+ \overline{\chi} (i \slashed{D} - \mchi) \chi 
\end{equation}
when $\chi$ is a Dirac fermion. For the Majorana case, $\chi$ is 
self-conjugate and its Lagrangian is multiplied by $1/2$. The SU(2) 
covariant derivative is $D_\mu = \partial_{\mu} - i g_2 A^C_\mu\, T^C$
where $T^C$, $C=1,2,3$, are the  SU(2) generators in the isospin-$j$ 
representation and $A_\mu^C$ are the EW gauge bosons. 
In these models the dark matter particle obtains the correct relic density 
from thermal freeze-out for $m_\chi$ in the 1-10 TeV range 
\cite{Cirelli:2005uq} for the favoured small representations $j=1,2$.

\subsection{Effective theory framework}
\label{sec:framework}

We consider the process
\begin{equation}
\chi^0(p_1)+ \chi^0(p_2) \to \gamma(p_\gamma) + X (p_X)
\end{equation}
for nearly maximal photon energy. Since the kinetic energy of the dark 
matter particles is negligible, $E^{\gamma}_{\rm max} = \mchi$. The 
theoretically calculated energy spectrum is distribution-valued. We 
integrate the energy spectrum over an interval $E^\gamma_{\rm res}$ 
from the endpoint, 
\begin{equation}
\langle \sigma v \rangle (E^\gamma_{\rm res}) = 
\int_{\mchi-E^\gamma_{\rm res}}^{\mchi} 
dE_\gamma\, \frac{d(\sigma v)}{dE_\gamma}\,.
\label{eq:defines}
\end{equation}
Roughly speaking, for a $\gamma$-telescope with resolution 
$E^\gamma_{\rm res}$, this together with the astrophysical line-of-sight 
factor determines the flux of photons from dark matter annihilation into  
the energy bin that contains the photon line signal. To turn experimental 
limits into model constraints, requires to replace the above integral 
by an integral with the experiment-specific response function. 
Eq.~(\ref{eq:defines}) is useful to quantify the importance of radiative 
corrections and resummation for the semi-inclusive energy spectrum 
from dark matter annihilation for $E_\gamma$ near $\mchi$, and 
will be used below.

The photon endpoint spectrum  
depends on four important scales: $\mchi$ (hard), the small invariant 
mass $m_X = \sqrt{4 m_\chi E^\gamma_{\rm res}}$ (collinear) of the 
unobserved, energetic final state, enforced by the kinematics of the 
endpoint, the electroweak scale $m_W$ (soft) and the energy resolution 
scale $E^\gamma_{\rm res}$ (ultrasoft). 
We shall now assume that the energy resolution is parametrically of order 
$E^\gamma_{\rm res} \sim m_W^2/\mchi$, which implies $m_X\sim m_W$ and 
the scale hierarchy $E^\gamma_{\rm res} \ll m_W,m_X \ll \mchi$. The 
factorization of the multi-scale Feynman diagrams into single-scale 
contributions, which is a prerequisite to all-order resummation, then  
requires the introduction of momentum modes with the following parametric 
scaling:
\begin{align}
\text{hard}\,(h):\quad & k^\mu \sim  \mchi (1,1,1)
\nonumber \\
\text{collinear}\,(c): \quad & k^\mu \sim  \mchi (1,\la^2,\la)
\nonumber \\
\text{anti-collinear}\,(\bar{c}): \quad & k^\mu \sim  \mchi (\la^2,1,\la)
\nonumber \\
\text{soft}\,(s): \quad & k^\mu \sim  \mchi (\la,\la,\la)
\\
\text{potential}\, (p) :\quad & k^0\sim m^2_W/\mchi,\, \mathbf{k}\sim m_W
\nonumber\\
\text{ultrasoft}\,(s): \quad & k^\mu \sim  \mchi (\la^2,\la^2,\la^2)
\nonumber
\end{align}
Here $\lambda = \sqrt{E^\gamma_{\rm res}/\mchi}$ and $k^\mu \sim 
(\np\cdot k,\nm\cdot k,\,\mathbf{k}_\perp)$ where $\np^\mu,\nm^\mu$ are 
two light-like vectors with $p_\gamma^\mu = E_\gamma \np^\mu$ and 
$\np\cdot\nm=2$. We remark that the collinear, anti-collinear, soft and 
potential modes all have the same virtuality ${\cal O}(m_W^2)$. The 
interactions of these modes are described by standard potential 
non-relativistic and soft-collinear effective Lagrangians (similar to 
\cite{Beneke:2010da}, generalized from QCD to the electroweak interaction).

The energy resolution of existing instruments is considerably 
larger than assumed here. For $\mchi = 3\,$TeV, allowing for 
$E_{\rm res}^\gamma \sim 4 m_W^2/m_\chi$ amounts to a resolution 
of about 10~GeV or $0.3\%$ compared to 10\% of, for example, H.E.S.S.
One might also consider the wider resolution $E^\gamma_{\rm res} \sim m_W$, 
which implies  $E^\gamma_{\rm res}, m_W \ll m_X \ll \mchi$ and a different 
mode structure. Conceptually, the main difference is caused by the fact 
that the previous, narrower resolution does not allow the radiation 
of soft particles with electroweak-scale masses into the unobserved final 
state. Although the resolution of the up-coming $\gamma$-ray telescopes 
is probably closer to the wide resolution case, in this work we concentrate 
on the narrow resolution $E^\gamma_{\rm res} \sim m_W^2/\mchi$ to stay 
close to the line-like signal. The wide resolution case, which is in 
fact simpler from the EFT point of view, will be discussed in subsequent 
work, which will also provide the explicit forms of the effective 
Lagrangians.

\subsection{Factorization}
\label{sec:factorization}

The primary annihilation process is described at leading order in an 
expansion in $\lambda$, which is also an expansion in $m_W/\mchi$, 
by operators ${\cal O}_i$ for the S-wave annihilation 
of the dark matter particles into two EW gauge bosons. Once the 
hard modes are integrated out into the coefficient functions $C_i$ of these 
operators, the collinear, anti-collinear and potential fields can no longer 
interact directly, since their momenta would add up to hard virtualities. 
The collinear modes build up the unobserved final state $X$, while the 
anti-collinear modes must result in the single, observed photon. The 
non-relativistic DM particles are described by the potential 
fields exchanging potential EW gauge bosons, which causes the 
Sommerfeld effect.

The factorization formula then follows from an analysis of the coupling 
of the (ultra) soft modes to any of the above. The decoupling of soft 
gauge boson attachments from heavy-particle fields in the presence 
of the Sommerfeld effect by a time-like Wilson-line field redefinition 
of the heavy-particle field has been demonstrated in \cite{Beneke:2010da} 
for an unbroken gauge theory. Although in the present case the Sommerfeld 
effect must be computed in the broken theory with gauge boson masses, 
soft attachments still factorize from the ladder diagrams, since a soft 
momentum throws the potential heavy particle off-shell, which removes 
the enhancement of the ladder rungs between the soft attachment and the 
hard vertex. It follows that the Sommerfeld factor $S_{IJ}$ completely 
factorizes from the Sudakov resummed annihilation rate 
$\Gamma_{IJ}(E_\gamma)$, 
\begin{equation}
\frac{d (\sigma v_{\text{rel}})}{d E_{\gamma}} 
= 2 \sum_{I,J} S_{IJ} \,\Gamma_{IJ}(E_\gamma)\,,
\label{eq:SIJGIJ}
\end{equation}
where the sums over $I,J$ run over all electrically neutral two-particle 
states that can be formed from the $2j+1$ single-particle states of the 
electroweak DM multiplet. For example, in the triplet (`wino') model, 
$I,J = \chi^0\chi^0, \chi^+\chi^-$. Since gauge boson exchange between 
the DM particles prior to annihilation can change the initial 
two-particle state $\chi^0\chi^0$ into any $I,J$, the annihilation rate 
with the Sommerfeld effect factored out 
is a matrix describing the amplitude for the 
annihilation of two-particle state $I$ times the complex conjugate 
of the annihilation amplitude for state $J$. The Sommerfeld factor 
is defined in terms of the matrix element of non-relativistic DM 
fields of the schematic form $\langle \chi^0\chi^0|[\chi\chi]_J|0\rangle 
\langle 0| [\chi\chi]_I | \chi^0\chi^0 \rangle$ such that 
$S_{IJ}=\delta_{IJ}$ in the absence of the potential force causing the 
Sommerfeld effect (see 
\cite{Beneke:2014gja} for the full definitions).
 
The coupling of soft gauge fields to collinear and anti-collinear fields is 
removed by a redefinition of the (anti) collinear fields with light-like 
Wilson lines \cite{Bauer:2001yt}. Since the small energy resolution forbids 
{\em soft} radiation into the final state $X$, the soft function is a
vacuum matrix element of Wilson lines that can be regarded as a soft 
Wilson coefficient $D$ of the annihilation 
{\em amplitude}~\cite{BSUunpublished,Urban:2017} on top of the hard 
Wilson coefficients $C_i$ of the operators ${\cal O}_i$. At this point 
all modes have been factored into separate functions except for the 
coupling of ultrasoft fields. At leading order in the expansion 
in $\lambda$, the coupling of ultrasoft gauge fields is again 
described by Wilson lines, where now the gauge field must be the photon. 
Since the initial state is electrically neutral and at rest, the 
coupling of ultrasoft photons is of higher-order, leaving ultrasoft Wilson 
lines from the (anti) collinear fields. There is no kinematic restriction on 
the radiation of ultrasoft particles with energy and masses of order 
$E_\gamma^{\rm res}$ into the final state, hence the ultrasoft function 
corresponds to an amplitude squared summed over the unobserved ultrasoft 
final state. 

We can therefore write down the following factorization formula for the 
energy spectrum of DM annihilation (more precisely, the off-diagonal 
annihilation matrix $\Gamma_{IJ}$ in the DM two-particle states) into a 
single identified-photon inclusive final state near the maximal photon 
energy:
\begin{eqnarray}
\Gamma_{IJ}(E_\gamma)&=& 
\frac{1}{(\sqrt{2})^{n_{id}}}\,
\frac{1}{4} \,\frac{2}{\pi\mchi} 
\sum_{i,j\,=\,1,2} \,\sum_{V,W,X,Y}
C^{*}_{j}(\mu_W)C_{i}(\mu_W)
D_{J,XY}^{j \, *}(\mu_W,\nu_s) D_{I,VW}^{i}(\mu_W,\nu_s)
\nonumber\\
&&\times \,V(\mu_W,\nu_s,\nu_j)\, Z_{\gamma}^{YW}\,
\int d\omega \, J^{XV}(4\mchi (\mchi-E_\gamma-\omega),\mu_W,\nu_j)
\, S_{\gamma}(\omega) 
\label{eq:factformula}
\end{eqnarray}
The prefactors account for the spin average, flux factor and photon 
momentum angular integration, $C$ and $D$ refer to the hard and soft 
matching coefficients of the annihilation amplitude.\footnote{The factor 
$1/(\sqrt{2})^{n_{id}}$ with $n_{id} = 0,1,2$ depending on how often the 
identical particle state $\chi^0\chi^0$ appears in $IJ$ arises, because 
we employ what was termed `method-2' in \cite{Beneke:2014gja} to evaluate 
the Sommerfeld effect. This is also the reason behind overall factor of 2 appearing in~\eqref{eq:SIJGIJ}.} The former is evolved 
from the scale $2\mchi$ to the electroweak scale $\mu_W$. The indices 
$V,W,X,Y$ refer to the SU(2) adjoint representation of the electroweak 
gauge bosons. In the second line $Z_\gamma$ is the anti-collinear factor 
for the observed photon, $J$ the jet function for the unobserved 
collinear final state $X$, convoluted with the ultrasoft function $S$.  
Since the (anti) collinear and soft modes have parametrically equal 
virtualities, a rapidity evolution factor $V(\mu_W,\nu_s,\nu_j)$ is needed. 
We postpone the derivation of this formula and technical details to a 
separate paper and proceed by defining the appearing functions and 
providing the results of their calculation.

\subsubsection{Operator basis and hard matching coefficients}

The relevant hard annihilation operators can be written as 
\begin{eqnarray}
\mathcal O_1 &=& \chi_v^{c\dagger}\Gamma^{\mu\nu}\chi_v\,
\mathcal{A}^B_{\perp c,\mu}(sn_+) \mathcal{A}^B_{\perp \bar{c},\nu}(tn_-)\,,
\label{eq:opbasis1}
\\
\mathcal O_2 &=& \frac{1}{2} \,\chi_v^{c\dagger}\Gamma^{\mu\nu}
\{T^B,T^C\}\chi_v\,
\mathcal{A}^B_{\perp c,\mu}(sn_+) \mathcal{A}^C_{\perp \bar{c},\nu}(tn_-)\,.
\label{eq:opbasis2}
\end{eqnarray}
Here $\chi_v$ is a two-component non-relativistic spinor field in the 
SU(2) weak isospin $j$ representation, $\chi_v^c = -i\sigma^2 \chi_v^*$ 
the charge-conjugated field, and $\mathcal{A}^B_{\perp c,\mu}$ the 
collinear gauge-invariant collinear gauge field of soft-collinear 
effective theory (SCET). Collinear fields have large momentum component 
$n_+p$. A similar definition applies to the anti-collinear direction 
with $n_+$ and $n_-$ interchanged. Fields without position arguments are 
evaluated at $x=0$. The operators are non-local along the light-cone 
of the collinear directions. $T^A$ are SU(2) generators in the isospin-$j$ 
representation. The spin matrix is defined by (conventions 
$n^\mu_\pm = (1,0,0,\mp 1)$, $\epsilon^{0123}=-1$)
\begin{equation}
\Gamma^{\mu\nu} = \frac{i}{4}\,[\sigma^\mu,\sigma^\nu] \,\sigma^\alpha 
(n_{-\alpha}-n_{+\alpha}) 
= \frac{1}{2 i}\,[\sigma^m,\sigma^n] \,\mathbf{\sigma}
\cdot\mathbf{n} \stackrel{d=4 \;\mbox{\tiny only}} 
= \frac{1}{2} \epsilon^{\mu\nu\alpha\beta} n_{+\alpha} n_{-\beta} 
\equiv \epsilon_{\perp}^{\mu\nu}\,.
\label{eq:spinmatrix}
\end{equation}
The first form holds in $d$ dimensions in dimensional regularization, but 
it turns out that evanescent operators are not important, since the 
non-relativistic, soft and collinear dynamics is spin-independent. 
$\mathbf{n}$ is a unit vector in the three-direction. Since the (anti) 
collinear gauge field in the operator is transverse, the Lorentz indices 
$\mu,\nu$ and corresponding spatial indices $m,n$ are effectively always 
transverse. Note that the DM bilinear is in a spin-singlet 
configuration. The above operator basis is consistent with the 
basis employed in~\cite{Ovanesyan:2014fwa,Ovanesyan:2016vkk}. 

We normalize the effective annihilation Lagrangian as 
\begin{equation} 
\frac{1}{2 m_\chi} \sum_{i=1,2} C_i(\mu) \mathcal{O}_i. 
\end{equation}
The one-loop calculation of the 
$\overline{\rm MS}$-subtracted matching coefficients results in 
\begin{eqnarray}
C_1(\mu_h) &=& \frac{\hat g^4_2(\mu_h)}{16 \pi^2}\, c_2(j)\, 
\Big[ (2-2 i \pi) \ln\Big(\frac{\mu_h^2}{4 \mchi^2}\Big) - 
\Big(4 - \frac{\pi^2}{2}\Big)\Big]  \,,
\label{eq:chard1}\\
C_2(\mu_h) &=& \hat g^2_2(\mu_h) + \frac{\hat g^4_2(\mu_h)}{16 \pi^2}
\Big[16 - \frac{\pi^2}{6} - c_2(j) 
\Big(10 - \frac{\pi^2}{2}\Big) - 6 \ln\Big(\frac{\mu_h^2}{4 \mchi^2}\Big) \, 
\nonumber \\
&&+ \,2 i \pi \ln\Big(\frac{\mu_h^2}{4 \mchi^2}\Big)
-2 \ln^2\Big(\frac{\mu_h^2}{4\mchi^2}\Big)\Big]  \,,
\label{eq:chard2}
\end{eqnarray}
where $\hat g_2(\mu_h)$ is the SU(2) gauge coupling in the $\overline{\rm MS}$ 
scheme at the matching scale $\mu_h \sim 2\mchi$, and $c_2(j) = j(j+1)$ the 
SU(2) Casimir of the isospin-$j$ representation.\footnote{For the wino 
model $j=1$, the one-loop coefficients were previously given in 
analytic form in~\cite{Ovanesyan:2016vkk} in the context of resumming the 
annihilation rate to the exclusive $\gamma\gamma$, $\gamma Z$ final state. 
When transforming to their operator basis, we find that our coefficient 
of $\hat g_2^4/(16 \pi^2)$ differs by +4 ($-4$) from their $C_1$ ($C_2$). 
We could track this difference to an error in the external field 
renormalization for the DM field and an inconsistency in combining 
counterterms for Dirac and Majorana $\chi$ fields. We note that the final 
result for the coefficients functions is independent of whether the DM 
particle is a Majorana or Dirac fermion.} 
The coefficients are evolved to the EW scale $\mu_W\ll \mu_h$. The 
evolution is diagonal~\cite{Beneke:2009rj} in the 
basis $\mathcal{O}_i^\prime$ where the DM bilinear transforms in an 
irreducible SU(2) representation given by 
\begin{equation}
\mathcal{O}^\prime = \hat{V}^T \mathcal{O},\qquad
\hat{V}=\left(\begin{array}{cc} 1 \;& -\frac{c_2(j)}{3} \\  
0 & \;1\end{array}\right),
\end{equation}
such that 
\begin{equation}
C(\mu_W) = \hat{V} 
\left(\begin{array}{cc} U^{(0)}(\mu_h,\mu_W) & 0 \\  
0 & U^{(2)}(\mu_h,\mu_W) \end{array}\right) 
\hat{V}^{-1}\,C(\mu_h)\,.
\end{equation}
The evolution factor in the irreducible isospin-$J$ representation satisfies 
the renormalization group equation
\begin{eqnarray}
\frac{d}{d\ln\mu} \,U^{(J)}(\mu_h,\mu) 
&=& \Bigg\{\,\frac{1}{2} \,\gamma_{\text{cusp}}\Bigg[2 c_2(\mbox{ad}) 
\Big(\ln\Big(\frac{4 \mchi^2}{\mu^2}\Big)-i \pi\Big)+i \pi c_2(J)\Bigg] 
\nonumber\\ 
&& + \, 2 \gamma_{\text{ad}} + \gamma^{J}_{H,s}\Bigg\}
\,U^{(J)}(\mu_h,\mu) 
\end{eqnarray}
with anomalous dimensions
\begin{eqnarray}
\gamma_{\text{cusp}}(\alpha_2) &=& \gamma_{\text{cusp}}^{(0)} 
\,\frac{\alpha_2}{4\pi} 
+ \gamma_{\text{cusp}}^{(1)} \left( \frac{\alpha_2}{4\pi} \right)^2 
+ \mathcal{O} (\alpha_2^3) \,, 
\\ 
\gamma^{(0)}_{\text{cusp}} &=& 4, \qquad
\gamma^{(1)}_{\text{cusp}} = \left(\frac{268}{9} - 
\frac{4 \pi^2}{3}\right)c_2(\text{ad}) - \frac{80}{9}n_G - \frac{16}{9}  \,,
\\
\gamma_{\text{ad}}(\alpha_2) &=& \gamma^{(0)}_{\text{ad}} 
\,\frac{\alpha_2}{4\pi} + \mathcal{O} (\alpha_2^2) \, , \\
\gamma^{(0)}_{\text{ad}} &=& - \beta_{0, \text{SU(2)}} = 
- \left( \frac{43}{6} - \frac{4}{3}n_G \right) ,
\\
\gamma^J_{H,s}(\alpha_2) &=& \gamma^{(0)}_{H,s}\, c_2(J) 
\,\frac{\alpha_2}{4\pi} + \mathcal{O} (\alpha_2^2) \, , \\
\gamma^{(0)}_{H,s} &=& -2 \,.
\end{eqnarray}
Here $c_2(\text{ad}) = 2$ is the Casimir for the adjoint representation 
and $n_G=3$ denotes the number of fermion generations. (The Higgs 
contribution $-16/9$ to the two-loop cusp anomalous dimension has been 
obtained from the $\epsilon$-scalar contribution in 
\cite{Broggio:2015dga}.) The NLL' and 
NLL approximations require the SU(2) cusp anomalous dimension in the SM 
in the two-loop approximation, and the other anomalous dimensions at 
the one-loop order, as given explicitly above. The LL approximation 
makes use only of the one-loop cusp term and neglects the other anomalous 
dimensions. 

\begin{figure}[t]
  \centering
  \includegraphics[width=0.505\textwidth]{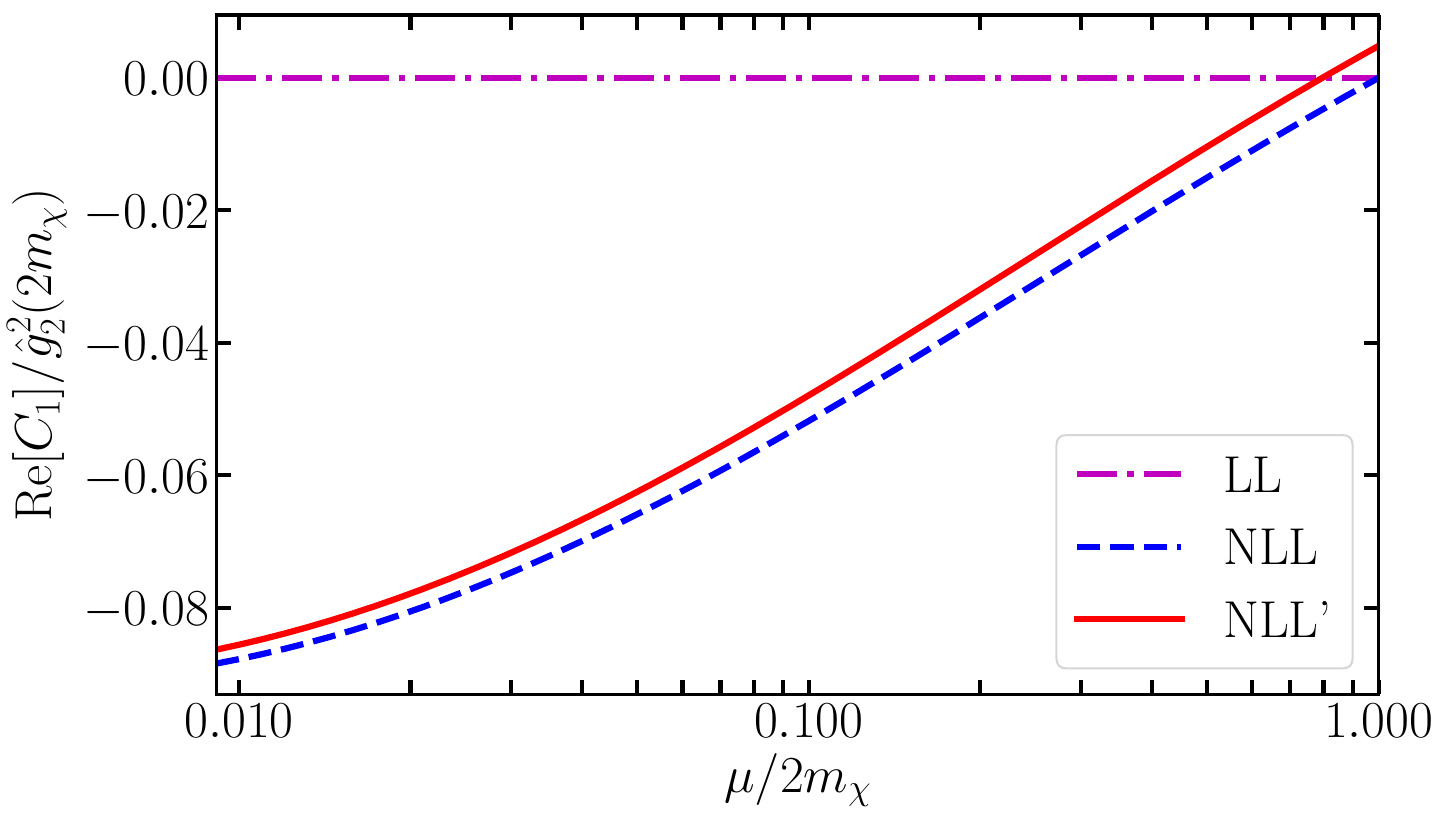} 
  \hspace*{0.1cm}
  \includegraphics[width=0.46\textwidth]{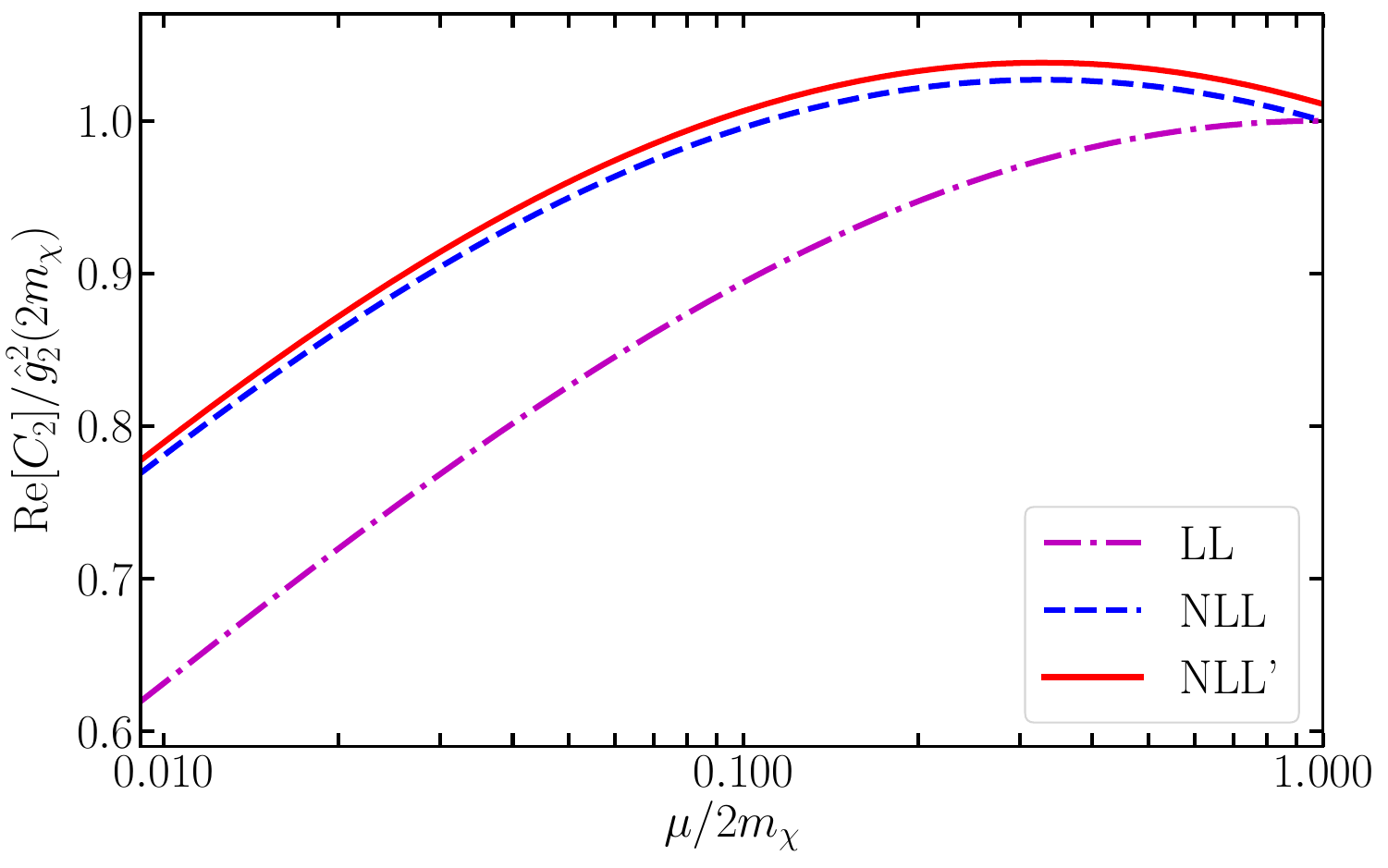}
\caption{Evolution of the real part of the matching coefficients in 
various approximations for $\mchi=5$~TeV, $\mu_h=2\mchi$.
\label{fig:cevolved}}
\end{figure}

In Figure~\ref{fig:cevolved} we show the evolved coefficient functions 
in the above mentioned approximations. The evolution equation is solved 
by numerical solution of the differential equation in the given 
approximation after solving the coupled system of renormalization group 
equations for the three gauge couplings, the top Yukawa and the Higgs 
self-coupling in the two-loop approximation. The input values for the 
couplings are specified at the scale $m_Z = 91.1876\,$GeV in the 
$\overline{\rm MS}$ scheme: 
$\hat \alpha_2(m_Z) = 0.0350009$, 
$\hat \alpha_3(m_Z) = 0.1181$, 
$\hat s_W^2(m_Z) = \hat g_1^2/(\hat g_1^2+\hat g_2^2)(m_Z) = 0.222958$,
$\hat \lambda_t(m_Z) = 0.952957$, 
$\lambda(m_Z) = 0.132944$. 
The $\overline{\rm MS}$ gauge couplings are computed via one-loop relations 
from $m_Z$, $m_W = 80.385\,$ GeV, the on-shell 
electromagnetic coupling $\alpha_{\text{OS}}(m_Z) = 1/128.943$ 
at the $Z$ mass scale, 
and the top Yukawa and Higgs self-coupling, which enter our calculation 
only implicitly through the two-loop evolution 
of the gauge couplings, via tree-level relations to 
$\overline{m}_t(\overline{m}_t)=163.35\,$GeV (corresponding to 
the top pole mass 173.2~GeV at four loops) and $m_H = 125.0\,$GeV. 

\subsubsection{Soft functions}

The soft renormalization factor of the annihilation vertex is given by 
the vacuum expectation value of the Wilson lines that arise from 
decoupling soft EW gauge bosons from the fields that appear in the 
operators ${\cal O}_i$. In (\ref{eq:factformula}) the soft factor 
$D^i_{I,VW}$ is defined in the basis of DM two-particle states with 
respect to the SU(2) indices of the DM bilinear, which corresponds to  
the definition
\begin{equation}
D^i_{I,VW} = K_{I,ab}\,
\langle 0|[Y_v^\dagger T_i^{AB} Y_v]_{ab} \,Y_{\nm}^{AV} 
Y_{\np}^{BW}|0\rangle\,.
\end{equation}
The matrix $K$ takes the linear combination appropriate to the 
two-particle state $I$, and $T_1^{AB} = \delta^{AB}$, $T_2^{AB} = 
\frac{1}{2}\,\{T^A,T^B\}$ for the two operators $i=1,2$. 
The light-like Wilson lines $Y_{\nm}^{AV}, Y_{\np}^{BW}$ arise from the 
gauge fields and are in the adjoint representation. The time-like 
Wilson line $Y_v$ is in the isospin-$j$ representation of the DM field.
All four Wilson lines extend from $x=0$ to infinity in their respective 
directions. Note that although $V,W=1,2,3$ are the gauge boson indices 
referring to $W^A$ rather than the mass eigenstates $W^\pm$, $Z$, $\gamma$, 
the soft function lives at the electroweak scale and must be computed 
with the Feynman rules of the SM after electroweak symmetry 
breaking, including gauge boson masses, contrary to the hard coefficient 
functions discussed above, which can be computed in the unbroken 
theory, neglecting the masses of the SM particles.

The requirement of an observed energetic photon implies $Y=W=3$, 
and then electric charge conservation implies that also $V=X=3$.
Hence, only the index values $V,W,X,Y=3$ contribute 
to (\ref{eq:factformula}), so the corresponding sums disappear. 
The NLL' approximation requires the one-loop calculation of every 
function that appears in the factorization formula. For the triplet 
(`wino') model we find  
\begin{eqnarray}
&& D^{1}_{(00),\,33}(\mu,\nu) = 1 + \frac{\hat g_2^2(\mu)}{16\pi^2} 
\Bigg(8 \ln^2 \frac{m_W}{\mu} -8 i \pi \ln\frac{m_W}{\mu} -\frac{\pi^2}{3}
-16 \ln\frac{m_W}{\mu} \ln \frac{m_W}{\nu}\Bigg)\,, \qquad\\[-0.1cm]
&& D^{2}_{(00),\,33}(\mu,\nu) = \frac{\hat g_2^2(\mu)}{16\pi^2} \,(8-8 i \pi) 
\ln \frac{m_W}{\mu}\,,\\[0.1cm]
&& D^{1}_{(+-),\,33}(\mu,\nu) = D^{1}_{(00),\,33}(\mu,\nu)\,, \\[0.1cm]
&& D^{2}_{(+-),\,33}(\mu,\nu) = D^{1}_{(00),\,33}(\mu,\nu) - \frac{1}{2}\,
D^{2}_{(00),\,33}(\mu,\nu)
\end{eqnarray}
for the two operators and the two distinct two-particle states $I=00,+-$. 
Since there are three regions with equal virtuality ${\cal O}(m_W^2)$ 
but different light-cone momentum component 
$\np\cdot k$, the soft function is not well defined with 
only a dimensional regulator. We use the rapidity regulator 
\cite{Chiu:2012ir} in addition to dimensional regularization to obtain 
the above result and the jet functions below. The scale $\nu$ is 
related to the rapidity regulator. The soft function contains no large 
logarithms if $\mu,\nu \sim {\cal O}(m_W)$. 

\subsubsection{Photon jet function}

The `jet' function for the exclusive anti-collinear photon state 
is defined by the squared matrix element 
\begin{equation}
-g_{\perp, \mu\nu} \,Z_\gamma^{BC} = 
\sum_{\lambda} \,
\langle 0|\mathcal A^B_{\perp \bar{c}, \mu}(0)|\gamma(p_\gamma,\lambda) 
\rangle\langle \gamma(p_\gamma,\lambda)|\mathcal A^C_{\perp \bar{c}, \nu}(0)|0\rangle
\end{equation}
of the three-component of the transverse SU(2) gauge field. 
$\mathcal A^B_{\perp \bar{c}, \mu}$ denotes the gauge field dressed with anti-collinear 
Wilson lines, $\hat{g}_2 \mathcal{A}^B_{\perp \bar{c}, \mu} T^B = 
W^\dagger_{\bar c} [i D_{\perp \bar{c}, \mu} W_{\bar c}]$, 
hence $Z^{33}_\gamma/\hat s_W^2$ can be interpreted as the on-shell photon 
field renormalization constant in light-cone gauge.

At the one-loop order we obtain 
\begin{eqnarray}
Z_\gamma^{33}(\mu,\nu) &=& \hat s_W^2(\mu) 
-\frac{\hat \alpha(\mu)}{4\pi}\,\bigg\{
- 16\ln\frac{m_W}{\mu}\ln\frac{2m_\chi}{\nu}
+8\ln \frac{m_W}{\mu}
\nonumber\\
&&   
- \,\hat s^2_W(\mu)\frac{80}{9}\bigg(\ln \frac{m_Z^2}{\mu^2}-\frac{5}{3}\bigg)
- \hat s^2_W(\mu)  \frac{16}{9} \ln \frac{m_t^2}{\mu^2}\nonumber\\
&& +\,\hat s^2_W(\mu)\bigg(3\ln \frac{m_W^2}{\mu^2} -\frac23\bigg) 
- 4\frac{m_W^2}{m_Z^2}\ln\frac{m_W^2}{\mu^2}\bigg\}
-\hat s_W^2(\mu)\Delta\alpha\,,
\end{eqnarray}
where $\Delta \alpha$ determines the difference between the fine structure 
constant $\alpha=1/137.036$ and 
$\alpha_{\rm OS}(m_Z) = \alpha/(1-\Delta \alpha)$.

\subsubsection{Jet function of the unobserved final state}

The jet function pertaining to the inclusive (unobserved) collinear 
final state is defined as the total discontinuity 
\begin{equation}
J^{BC}(p^2) = \frac{1}{\pi}\,\text{Im}\big[\,
i\mathcal{J}^{BC}(p^2) \big]
\end{equation}
of the gauge-boson two-point function 
\begin{equation}
-g_{\mu \nu} \,\mathcal{J}^{BC}(p^2)\equiv 
\int d^4x \,e^{ip\cdot x} 
\langle 0|\mathbf{T}\big\{\mathcal{A}^B_{\perp c, \mu}(x) 
\,\mathcal{A}_{\perp c, \nu}^C(0)\big\} |0\rangle\,.
\end{equation}
Again the field $\mathcal{A}^B_{\perp c, \mu}$ refers to the 
collinear gauge-invariant gauge field, which equals the ordinary 
gauge field in light-cone gauge. 

\begin{figure}[t]
  \centering
  \includegraphics[width=0.8\textwidth]{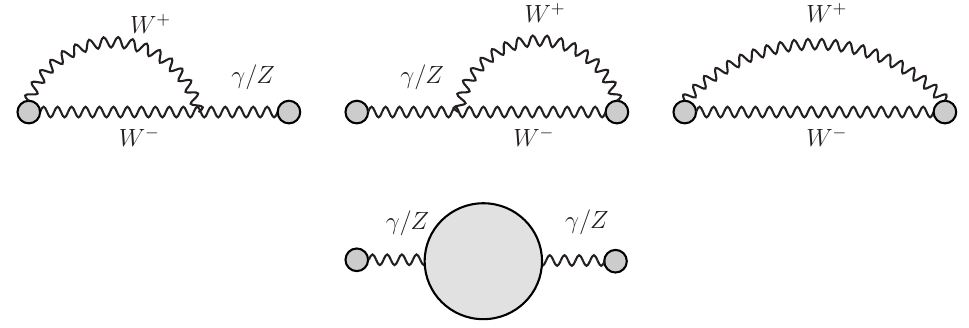} 
\caption{Wilson-line and self-energy type one-loop contributions to the 
jet function. 
\label{fig:jetfunctiondiagrams}}
\end{figure}
We compute the 33 component in Feynman gauge to the one-loop order 
and write $J^{33}(p^2)$ in the form 
\begin{equation}
J^{33}(p^2) = \hat s_W^2(\mu)\,\delta(p^2)
+\hat c_W^2(\mu)\,\delta(p^2-m_Z^2) 
+ J^{33}_{\text{Wilson line}}(p^2) + J^{33}_{\text{se}}(p^2)\,.
\end{equation}
The one-loop correction is split into two contributions, of which 
the first refers to diagrams that involve at least one contraction 
with a gauge field from a Wilson line in the definition of 
$\mathcal{A}^B_{\perp c, \mu}$ and the second to the remaining diagrams, 
which are of self-energy type, as shown in the first and second 
line of Fig.~\ref{fig:jetfunctiondiagrams}, respectively. 
Only the Wilson-line diagrams require the rapidity regulator, and 
their sum is given by 
\begin{eqnarray}
J^{33}_{\text{Wilson line}}(p^2,\mu,\nu) &=& 
 - \frac{\hat s^{2}_W(\mu) \hat{g}^2_2(\mu)}{16 \pi^2} \bigg\{ 
\delta(p^2)\bigg[ - 16\ln \frac{m_W}{\mu} \ln\frac{2 m_{\chi}}{\nu}
 + 8\ln\frac{m_W}{\mu}\bigg] 
\nonumber\\
&& \hspace*{-2cm} 
+ \,\frac{1}{p^2}\,\theta(p^2-4 m^2_W)\Big[ 4 \beta 
+ 8\ln \frac{1-\beta}{1+\beta}\,\Big]\bigg\}
\nonumber \\
&& \hspace*{-2cm} 
- \,\frac{\hat c^{2}_W(\mu) \hat g^2_2(\mu)}{16 \pi^2} \bigg\{ 
\delta(p^2-m^2_Z) \bigg[ - 16\ln\frac{m_W}{\mu}\ln\frac{2m_\chi}{\nu}
+8\ln\frac{m_W}{\mu}- 8 + 4\pi^2 
\nonumber\\
&&\hspace*{-2cm}
+\, 4\pi \bar{\beta}_Z -(16\pi +8 \bar{\beta}_Z) \arctan(\bar{\beta}_Z)
+ 16 \arctan^2(\bar{\beta}_Z)\bigg] 
\nonumber \\
&&\hspace*{-2cm}
+\, \frac{1}{p^2-m^2_Z}\,\theta(p^2-4 m^2_W)
\Big[ 4 \beta + 8\ln \frac{1-\beta}{1+\beta}\,\Big]\bigg\} \, ,
\end{eqnarray}
where 
\begin{equation}
\beta = \sqrt{1-\frac{4m^2_W}{p^2}}\,,\qquad
\bar{\beta}_Z = \sqrt{\,\frac{4m^2_W}{m^2_Z}-1} \,. 
\end{equation}
The self-energy type contribution $J^{33}_{\text{se}}(p^2)$ 
can be expressed in terms of 
conventional one-loop gauge-boson self-energies, which can be found, 
for example in \cite{Denner:1991kt}. 
We have taken the massless-quark limit of these 
expressions except for the top quark. We will provide the lengthy 
expressions in the detailed write-up. 

\begin{figure}[t]
  \centering
  \includegraphics[width=0.69\textwidth]{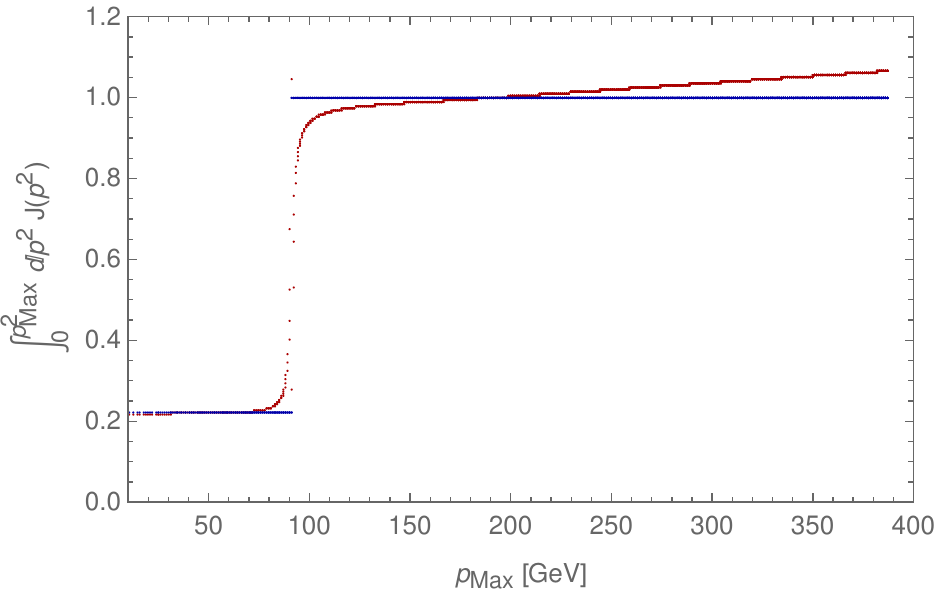} 
\vskip0.3cm
  \hspace*{-0.5cm}
  \includegraphics[width=0.727\textwidth]{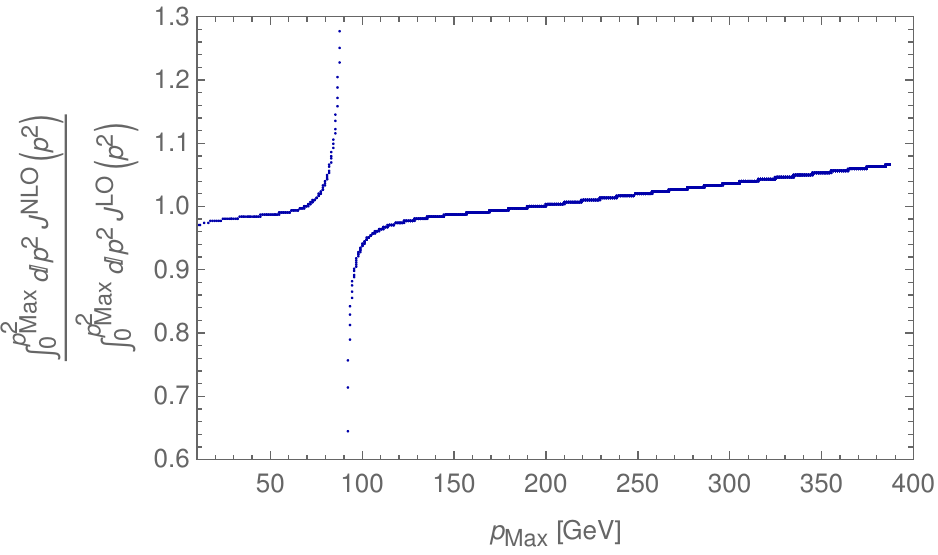}
\caption{Integrated jet function (upper panel) at next-to-leading order 
for $\mu=\nu=m_W$ and $\mchi = 2\,$TeV. The straight lines represent the 
leading-order result. The lower panel shows the 
integrated jet function at NLO normalized to its LO value.
\label{fig:jetfunction}}
\end{figure}

In Fig.~\ref{fig:jetfunction}
we show the dependence of the integrated jet function on the 
invariant mass of the unobserved collinear final state. The 
integrated function jumps from a value around $\hat{s}_W^2(m_W)$ 
to a value around 1 as the invariant mass passes through $m_Z$, and 
then slowly increases. The range shown contains the $W^+ W^-$, 
$ZH$ and $t\bar t$ thresholds ($m_t=173.2$~GeV is used here), 
which, however, are barely visible. The singularity of the NLO correction 
near $m_Z$ can be removed by a proper treatment of $Z$ boson resonance.
However, below we will adopt values $E^\gamma_{\rm res} > m_Z^2/(4 \mchi)$, 
which implies that $p^2_{\rm max}$ is always larger than $m_Z^2$. 

The jet functions contain no large logarithms when $\mu ={\cal O}(m_W)$ 
and $\nu = {\cal O}(\mchi)$. The `rapidity logarithms' related to the 
different value of $\nu$ that minimizes the logarithms in the soft and 
jet functions, respectively, can be summed at NLL' by solving the rapidity 
renormalization group equations \cite{Chiu:2012ir} in the one-loop 
approximation. For the case at hand we find 
\begin{equation}
V(\mu_W,\nu_s,\nu_j) = 
\exp \left[\frac{4c_2(\text{ad})}{\beta_{0,\rm SU(2)} }
\ln \left(\frac{\hat\alpha_2(\mu_W)}{\hat\alpha_2(m_W)}\right) 
\ln \frac{\nu_j^2}{\nu_s^2}\,\right].
\end{equation}
We checked that in the sum of all contributions the poles in the 
dimensional and rapidity regulator cancel. The hard, soft and jet 
functions above are defined by minimally subtracting the poles.

\subsubsection{Ultrasoft function}

The kinematics of the process does not allow soft radiation with 
momentum of order $m_W$ into the final state, which prohibits EW  
gauge boson radiation. However, radiation of photons and light 
quarks with masses of order or less than $m_W^2/\mchi$ is possible, 
which implies the convolution of the unobserved-final state jet function 
with an ultrasoft function accounting for the energy taken away from 
the collinear final state by ultrasoft radiation. 
The ultrasoft function is defined in terms of Wilson lines of ultrasoft 
photons and depends on the electric charges and directions 
of the particles in the initial and final state. After factoring the 
Sommerfeld effect, also the $\chi^+\chi^-$ initial state must be 
considered. But for the S-wave annihilation 
operators ${\cal O}_i$ only the total charge of the initial state 
is relevant for ultrasoft radiation, which vanishes. Furthermore, 
only the electrically neutral 33 components of the (anti) collinear 
functions appear for the $\gamma+X$ final state. We therefore conclude 
that the ultrasoft function is trivial,  
$S_\gamma(\omega) = \delta(\omega)\,.$
For this reason did not indicate the ultrasoft scale dependence of the 
functions in (\ref{eq:factformula}), and the convolution integral 
in (\ref{eq:factformula}) disappears. 

\subsubsection{Sommerfeld factor}

The various functions discussed above are assembled according to 
(\ref{eq:factformula}) into the annihilation matrix $\Gamma_{IJ}$ of the 
$\chi^0\chi^0$ and $\chi^+\chi^-$ DM two-particle states. In the 
process we checked the consistency of (\ref{eq:factformula}) by 
verifying that after renormalizing the parameters in the tree-level 
annhilation rates, all $1/\epsilon$ and rapidity regulator poles that 
appear in the various factors of the factorization formula cancel 
among each other, leaving logarithms consistent with the anomalous 
dimensions of these factors. The 
photon energy spectrum is then obtained according to 
(\ref{eq:SIJGIJ}) by tracing this matrix with the Sommerfeld 
factor $S_{IJ}$. For the fermionic DM triplet model, the Sommerfeld 
factors were first computed in \cite{Hisano:2004ds}. In the present 
work we employ the modified variable phase method \cite{Beneke:2014gja} 
for solving the Schr\"odinger equation and, different from the 
above, use the on-shell value $\alpha_2=0.0347935$ 
of the SU(2) coupling as well as $\alpha_{\rm OS}(m_Z)$  in the 
Yukawa-Coulomb potential for the two-state system. Near the 
Sommerfeld resonance, the result is very sensitive to the 
mass splitting between the electrically charged and neutral members  
of the triplet. The mass splitting with two-loop accuracy 
can be inferred from \cite{Yamada:2009ve,Ibe:2012sx}, and is given 
after adjustment to our input coupling parameters by
\begin{align}
\delta m_{\chi} = 164.1 \,\text{MeV}. 
\end{align}
The first and second Sommerfeld resonances are located at 
2.285~TeV and 8.817~TeV, respectively, for the coupling parameters and mass 
splitting employed in this work.

\section{Results}
\label{sec:results}

It is straightforward to calculate the one-dimensional integral
(\ref{eq:defines}) that defines the $\gamma+X$ yield from DM pair 
annihilation in a photon energy bin of size $E^\gamma_{\rm res}$. 
For the discussion below we shall assume $4 \mchi E^\gamma_{\rm res} 
=(300\,\text{GeV})^2$, which implies that the unobserved final 
state includes $\gamma$, $Z$, $W^+ W^-$, $Z H$ and light fermion 
pairs in the collinear jet function. 

Figure~4 shows (upper panel) our results for 
$\langle \sigma v\rangle(E^\gamma_{\rm res})$ 
as defined in (\ref{eq:defines}). 
The displayed DM mass range includes the 
first two Sommerfeld resonances. The four lines refer to the 
Sommerfeld-only calculation, which employs the tree-level 
approximation to $\Gamma_{IJ}$ (black-dotted) and the successive 
LL (magenta-dashed), NLL (blue-dashed) and NLL' (red-solid) resummed 
expressions for the same quantities with the latter representing 
the best approximation. 

The importance of resummation of electroweak Sudakov logarithms  
becomes more apparent by normalizing to the Sommerfeld-only 
result (lower panel). As expected and already seen in previous LL and NLL 
calculations \cite{Baumgart:2014vma,Bauer:2014ula,Baumgart:2015bpa,Ovanesyan:2014fwa,Ovanesyan:2016vkk} of related exclusive and semi-inclusive 
final states, resummation reduces the annihilation rate and increasingly 
so at larger DM mass. In the interesting mass range around 3~TeV where wino 
DM accounts for the observed relic density, the rate is suppressed 
by about a factor of two. 

\begin{figure}[t]
  \centering
  \hspace*{-0.6cm}
  \includegraphics[width=0.80\textwidth]{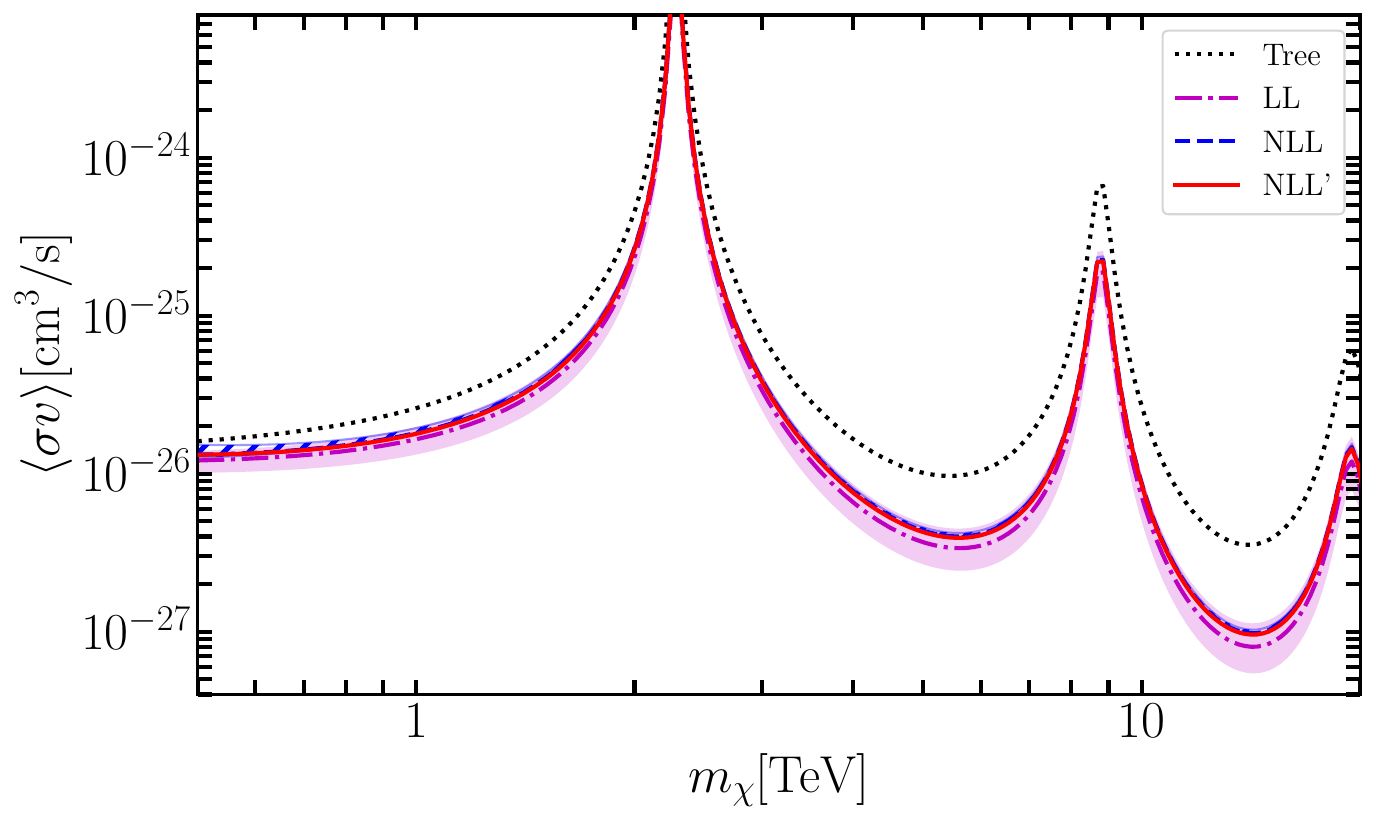} 
  \vskip0.2cm
  \includegraphics[width=0.77\textwidth]{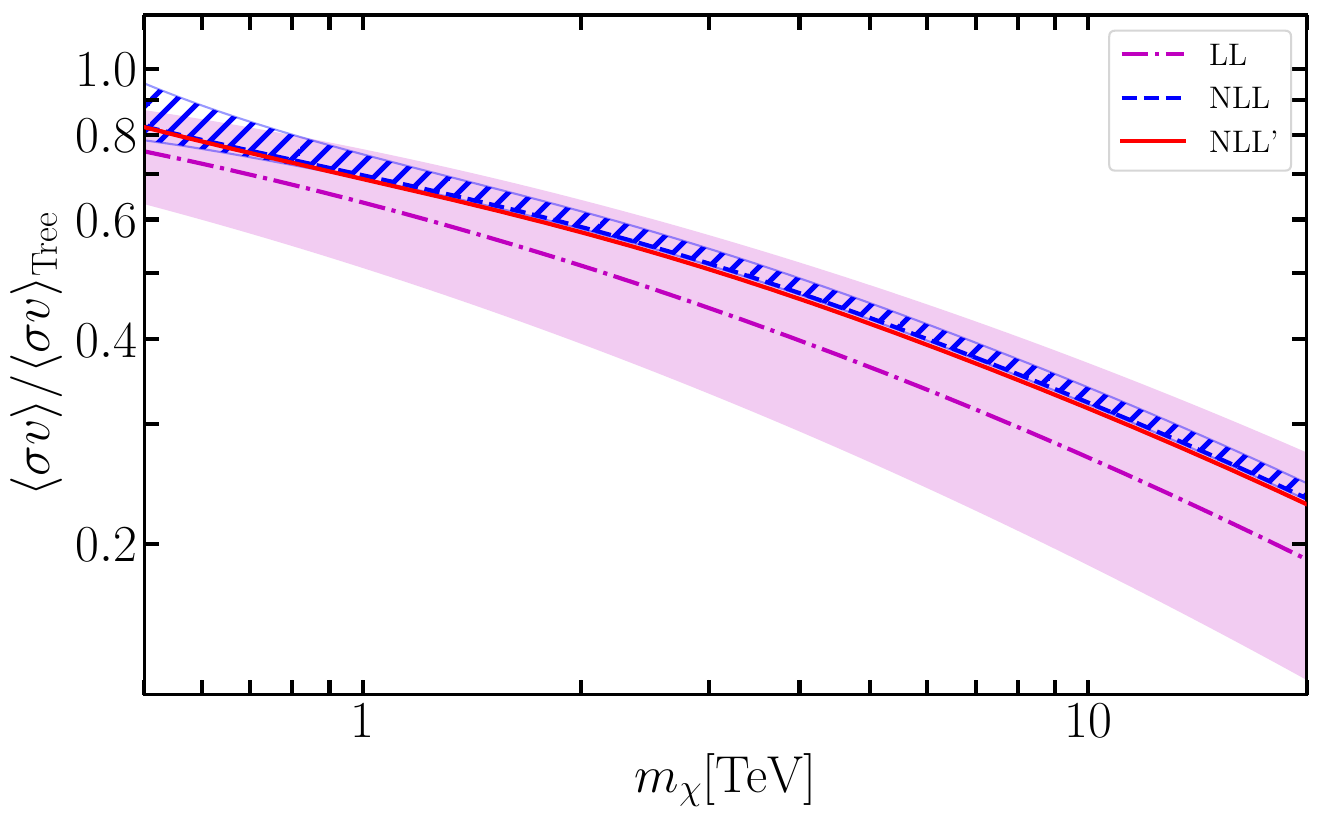}
\caption{Integrated photon energy spectrum within $E^\gamma_{\text{res}}$ 
from the endpoint $\mchi$ in the tree (Sommerfeld only) and LL, NLL, NLL' 
resummed approximation.
The shaded/hatched bands show the scale variation of the respective 
approximation as described in the text. For the NLL' result the theoretical 
uncertainty is given by the thickness of the red line.
\label{fig:result}}
\end{figure}

The resummed predictions are shown with theoretical uncertainty 
bands computed from the separate variations of the scales 
$\mu_h$, $\nu_j$ in the interval $[\mchi,4\mchi]$ and 
the scales $\mu_W$, $\nu_s$ in $[m_W/2,2 m_W]$, added in quadrature.  
We find that at the NLL' order the residual theoretical uncertainty 
from scale dependence is negligible -- in the figure it is 
given by the width of the red line.
Our result can be compared most directly with the recent work 
by Baumgart et al. \cite{Baumgart:2017nsr} who considered the same 
$\gamma+X$ final state at larger resolution $E_{\rm res}^\gamma \gg m_W$ 
with LL accuracy. We observe that the
inclusion of one-loop corrections to the hard, soft and jet functions
in our NLL' computation has the main effect of eliminating
the theoretical uncertainty of resummation by reducing the scale
dependence from 24\% (LL) to 3\% (NLL) to 0.3\% at NLL' at
$\mchi=2~$TeV. A similar reduction of scale dependence was already observed
in the NLL' calculation of the exclusive $\gamma\gamma$, $\gamma Z$
final state \cite{Ovanesyan:2016vkk}. In numbers we find that
at $\mchi = 2\,$TeV (10~TeV), the ratio of the resummed to the 
Sommerfeld-only rate is $0.513^{+0.128}_{-0.120}$ 
($0.268^{+0.101}_{-0.082}$) at LL, 
$0.585^{+0.032}_{-0.004}$ ($0.323^{+0.017}_{-0.002}$) at NLL and
$0.575^{+0.003}_{-0.000}$ ($0.316^{+0.002}_{-0.000}$) at NLL'. The 
central values are evaluated at the central scales 
of the intervals above. We also varied all four scales simultaneously 
within these intervals and determined the maximal variation. The 
scale dependence at NLL' with this more conservative procedure increases by 
about a factor of two, which does not change the general picture.


In conclusion, we computed the $\gamma+X$ spectrum near maximal photon 
energy from electroweak triplet (`wino') DM annihilation including the 
resummation of the Sommerfeld effect and electroweak Sudakov logarithms in 
the NLL' order. The inclusion of the electroweak one-loop corrections 
at NLL' renders the theoretical uncertainty of resummation 
negligible. It is plausible that the dominant theoretical uncertainty 
now arises from the fact that the non-relativistic EFT is only 
employed at leading order in the computation of the Sommerfeld 
enhancement, and from $\mathcal{O}(m_W/\mchi)$ power corrections. 
In a subsequent paper we shall present an extension of 
this work to the case of wider photon resolution, further details 
on the EFT framework, the derivation of the factorization 
formula, and a comparison with expected experimental limits. The 
computations performed here are presently restricted to simple DM 
models, which add to the SM a single electroweak multiplet. It 
would be of interest to extend them to more complex models such as 
the MSSM, which would put the analysis of indirect detection constraints 
for mixed DM models \cite{Beneke:2016jpw} on the same theoretical 
footing as for minimal models.

\subsubsection*{Acknowledgement}

We thank D. Pagani and K. Urban for discussions.
This work has been supported in part by the 
Collaborative Research Center `Neutrinos and Dark Matter in Astro- and 
Particle Physics' (SFB 1258) and the Excellence Cluster `Universe' 
of the Deutsche Forschungsgemeinschaft.  
MB thanks the Albert Einstein Institute at Bern University and 
the Department of Energy's Institute for Nuclear Theory at  the 
University of Washington for hospitality during the preparation 
of this work.


\providecommand{\href}[2]{#2}\begingroup\raggedright\endgroup


\end{document}